\documentclass[twocolumn,prl,secnumarabic,superscriptaddress]{revtex4}
\usepackage{amsfonts}
\usepackage{amsmath}
\usepackage{amssymb}
\usepackage{graphicx}
\usepackage{epstopdf}
\usepackage{float}
\begin{document}

\title{Spin-orbit anisotropy measured using ballistic spin resonance}

\author{W. W. Yu}
\affiliation{Department of Physics and Astronomy, University of British Columbia, Vancouver, British Columbia V6T 1Z4, Canada}
\author{S. M. Frolov}
\affiliation{Department of Physics and Astronomy, University of British Columbia, Vancouver, British Columbia V6T 1Z4, Canada}
\affiliation{Kavli Institute of Nanoscience, Delft University of Technology, 2600GA Delft, The Netherlands}
\author{S. L\"{u}scher}
\affiliation{Department of Physics and Astronomy, University of British Columbia, Vancouver, British Columbia V6T 1Z4, Canada}
\author{J. A. Folk}
\affiliation{Department of Physics and Astronomy, University of British Columbia, Vancouver, British Columbia V6T 1Z4, Canada}
\author{W. Wegscheider}
\affiliation{Laboratorium f\"{u}r Festk\"{o}rperphysik, ETH Z\"{u}rich, 8093 Z\"{u}rich, Switzerland}

\newcommand{\y}{$[\bar{1}10]$}
\newcommand{\x}{$[110]$}
\newcommand{\vnl}{$V_{nl}$}
\newcommand{\lsx}{\lambda_{[110]}}
\newcommand{\lsy}{\lambda_{[\bar{1}10]}}

\date{\today}
\begin{abstract}
Spin relaxation can be greatly enhanced in narrow channels of two-dimensional electron gas due to ballistic spin resonance, which is mediated by spin-orbit interaction for trajectories that bounce rapidly between channel walls. The channel orientation determines which momenta affect the relaxation process, so comparing relaxation for two orientations provides a direct determination of spin-orbit anisotropy.  Electrical measurements of pure spin currents are shown to reveal an order of magnitude stronger relaxation for channels fabricated along the \x\ crystal axis in a GaAs electron gas compared to \y\ channels, believed to result from interference between structural and bulk inversion asymmetries.
\end{abstract}

\maketitle

Spin-orbit interaction (SOI) is one of the most promising tools for fast spin control in solid state systems \cite{ZuticRMP04}.  Although SOI was first studied in the context of spin relaxation and decoherence, a variety of recent proposals and experiments have explored the use of SOI for intentional spin separation, filtering and coherent control in spintronic devices and spin qubits \cite{DattaAPL90, KatoScience04,GanichevNature06,KiselevAPL01,KogaPRL02,NowackScience07}.  Two-dimensional electron gases (2DEGs) make an attractive platform for spin-based electronics because electron transport can be easily controlled using electrostatic gates.

\begin{figure}[!h]
  \includegraphics[scale=1]{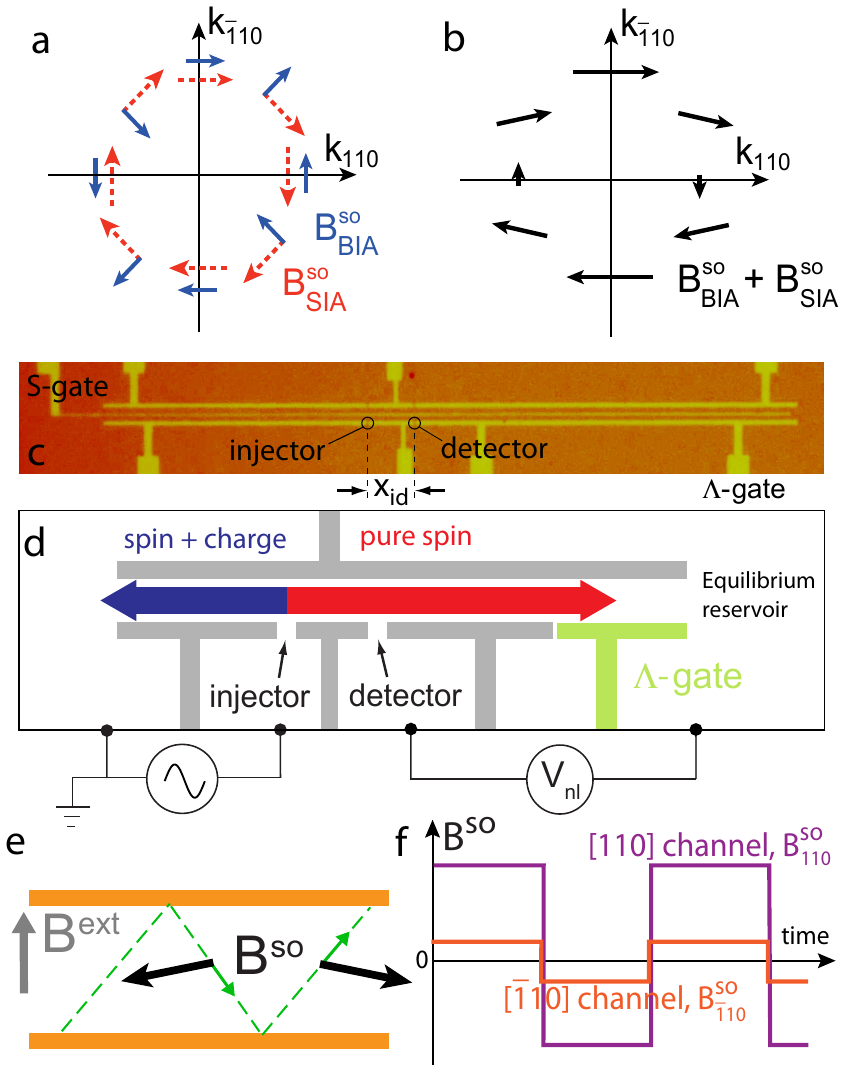}
  \caption{\textbf{(a, b)} Spin-orbit fields due to SIA and BIA and the total spin-orbit field, for the case $\alpha = -1.5\beta$.  \textbf{(c)} Optical image of a typical device (short channel). In the measurement, S-gate is depleted to form two 1 $\mu$m-wide channels from this single device. \textbf{(d)} Schematic of measurement setup. The channel and the injector and detector QPCs are defined by depleting the 2DEG with gates (light gray). $\Lambda$-gate changes the distance between the equilibrium reservoir and the injector.  \textbf{(e)} Periodic trajectories of electrons (dashed line and thin arrows) between channel boundaries are accompanied by an oscillating $B^{so}$ (thick black arrows). Injected spins are aligned transverse to the channel by the in-plane field $B^{ext}$. \textbf{(f)} Component of spin-orbit field along the channel axis oscillates with different magnitudes for \x\ and \y\ channels.}
 	\label{device}
\end{figure}

For 2DEGs in zinc-blende semiconductors such as GaAs, the primary contributions to SOI are structural inversion asymmetry (SIA) and bulk inversion asymmetry (BIA),  which add to induce an effective magnetic field, $\vec{B}^{so}$, that acts on the electron spin \cite{RashbaSPSS60, DresselhausPR55}:
\begin{equation}
\label{eq:bso}
\vec{B}^{so}=\frac{2}{g\mu_B}((\alpha-\beta)k_{\bar{1}10}\hat{\imath}-(\alpha+\beta)k_{110}\hat{\jmath})
\end{equation}
to first order in the momentum $k$, where $\hat{\imath}$ and $\hat{\jmath}$ are unit vectors along the \x\ and \y\ crystal axes, and $\alpha$ and $\beta$ represent the strength of the first-order SIA and BIA terms, respectively.   SIA and BIA vary from 2DEG to 2DEG depending on sample details, but $\alpha$ and $\beta$ are believed to be of the same order of magnitude in typical GaAs heterostructures (triangular wells). By Eq.~1, this should lead to anisotropy in the spin-orbit field, as the effects of SIA and BIA add to or subtract from one another depending on the direction of the electron momentum [see Fig.~\ref{device}(a)].   When $\alpha$ and $\beta$ have opposite signs, the interference between SIA and BIA gives rise to $\vec{B}^{so}$ that is stronger for momenta in the \y\ direction [Fig. \ref{device}(b)]; this anisotropy is reversed if $\alpha$ and $\beta$ have the same sign. 

Measuring spin-orbit anisotropy is crucial to the design of electronic devices for spin manipulation in a 2DEG, and for optimizing heterostructure growth for spintronics. The condition $|\alpha|=|\beta|$ is of great interest for applications where one seeks to rotate electron spin using SOI with d.c.~voltages \cite{SchliemannPRL03, CartoixaAPL03, OhnoPRB08, DyakonovSPSS72, KoralekNature09}.  Conversely, spintronics proposals based on the mesoscopic spin Hall effect depend on moving to $|\alpha|\ll|\beta|$ or $|\alpha|\gg|\beta|$\cite{SinitsynPRB04}.
Despite the importance of spin-orbit anisotropy to 2DEG spintronics, there are only a few published reports of $\alpha$ and $\beta$ in the triangular well heterostructures typically used for electronic devices, and the values found in these reports are often inconsistent\cite{GiglbergerPRB07, DenegaPRB10, MillerPRL03}.  The only transport measurement of spin-orbit parameters to date was based on indirect technique in which the shapes of magnetoresistance curves were compared to a detailed theory of weak antilocalization\cite{MillerPRL03}.

In this Letter, we report an all-electrical technique for directly accessing spin-orbit anisotropy in a GaAs/AlGaAs 2DEG. The method is based on ballistic spin resonance (BSR), which suppresses spin relaxation length in narrow conducting channels \cite{FrolovNature09}.  Spin-orbit anisotropy is measured by comparing BSR in 1 $\mu$m-wide 2DEG channels fabricated along the \x\ and \y\ crystal axes. The spin relaxation length at resonance is 3 $\mu$m in \x\ channels, compared to 40 $\mu$m in \y\ channels. Based on Monte-Carlo simulations of semiclassical spin dynamics, this anisotropy in relaxation length implies a spin-orbit anisotropy $|\alpha-\beta|/|\alpha+\beta|=15\pm$5, the largest value reported so far for GaAs triangular wells \cite{Luescher10}. Such a strong anisotropy will play an important role in the design of spin qubits and spin transistors in similar materials.

The channels are defined by electrostatic gates on the surface of a [001] GaAs/AlGaAs heterostructure.  The electron gas ($n_s=1.11 \times 10^{11}$ cm$^{-2}$ and mobility $\mu = 4.44 \times 10^6$ cm$^2$/Vs at 1.5K) is 75 nm below a layer of Si-doped AlGaAs, and 110 nm below the wafer surface. The data presented in this paper are from four devices: short and long channels (length 100 and 130 $\mu$m) aligned along \x\ and \y.  In short channels, the spacing between injector and detector contacts is $x_{id}$ = 5 $\mu$m; in long channels, $x_{id}$ = 25 $\mu$m.  Measurements are performed with an a.c. lock-in technique in a dilution refrigerator, with an external magnetic field $B^{ext}$ applied in the plane of the electron gas.

Two quantum point contacts (QPCs) are embedded into each channel.  At high magnetic field the QPC conductance traces show spin-polarized conductance plateaus at 1 $e^2/h$, which are used to generate and detect electron spins [see Figs.~\ref{device}(c) and \ref{device}(d)] \cite{vanweesprl88,WharamJPCSSP88,FrolovPRL09}.  The injector QPC drives polarized electrons towards the left end of the channel. Diffusion of spin polarization towards the equilibrium reservoir on the right end creates a pure spin current. Spin polarization accumulated above the detector QPC is then quantified by the non-local spin signal $V_{nl}$.

Electron trajectories in these devices bounce back and forth many times between the channel walls before scattering, because the mean free path is much longer than the channel width.  Ballistic spin resonance is driven by this motion, and mediated by spin-orbit coupling. As shown in Fig.~\ref{device}(e), an injected electron bouncing between the channel walls feels a spin-orbit field $B^{so}$, and the component of $B^{so}$ along the channel axis oscillates with the bouncing frequency [Eq.~\ref{eq:bso}].

The magnetic field $B^{ext}$ sets the spin quantization direction and the polarization of injected spins but does not affect the orbital motion of the electrons because it is applied in the plane of the electron gas. BSR is observed when $B^{ext}$ is applied perpendicular to the channel axis. In this case, the periodic component of $B^{so}$ is transverse to the external field, and it induces spin rotations when the Larmor precession frequency $g\mu B^{ext}/ h$ matches the typical bouncing frequency $v_F/2w$, where $v_F$ is the Fermi velocity and $w$ is the width of the channel. Spin polarization is rapidly lost on resonance because every electron follows a different bouncing trajectory.

\begin{figure}[!h]
  \includegraphics[scale=1]{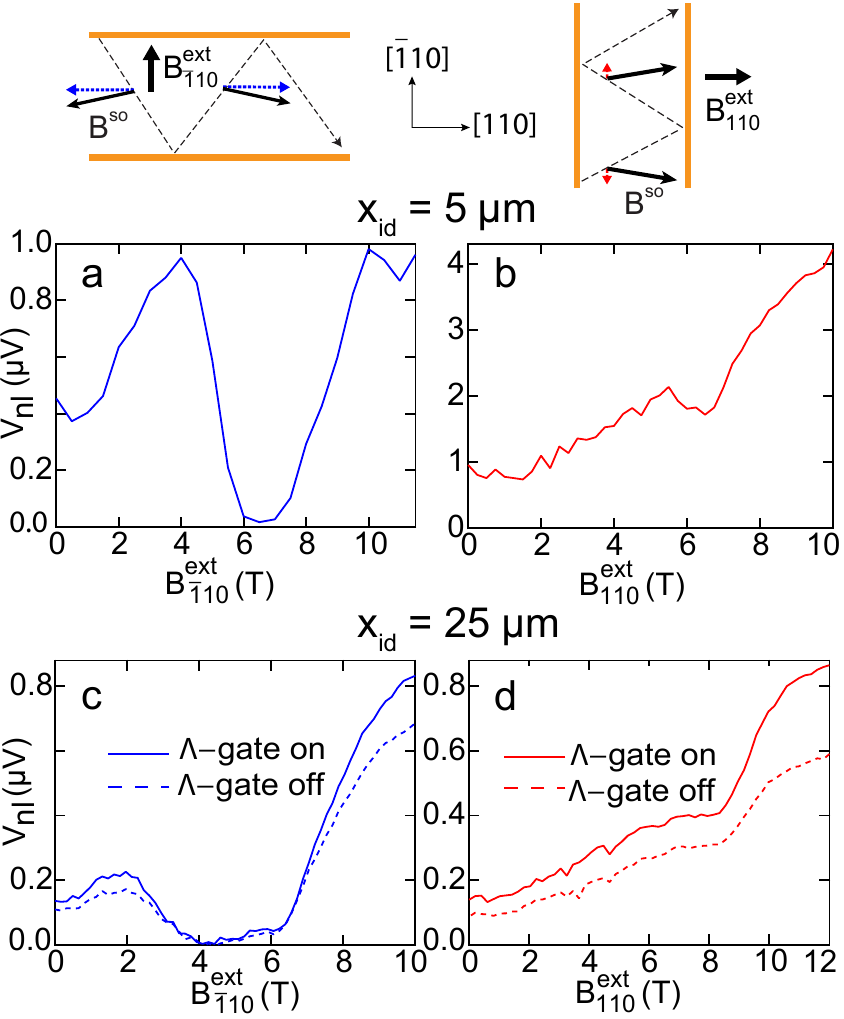}
  \caption{Magnetic field dependence of \vnl\ measured with injector and detector set to 1$e^2/h$ and separated by 5 $\mu$m \textbf{(a, b)} and 25 $\mu$m \textbf{(c, d)}.  Channel orientations are shown at the top: along [110] for \textbf{(a, c)} and along $[\bar{1}10]$ for \textbf{(b, d)}.  In \textbf{(c)} and \textbf{(d)}, dashed lines are \vnl\ measured with $\Lambda$-gate undepleted [see Fig.~\ref{device}(d)]. Finite signal at zero field in all panels is a signature of the Peltier effect, not of spin current \cite{FrolovPRL09}.}
 	\label{by}
\end{figure}

Figure 2 shows characteristic traces of $V_{nl}(B^{ext})$ for four different channels.  Two qualitative features are observed in all data sets.  First, $V_{nl}$ starts near zero and generally grows with magnetic field, showing the rise in QPC polarization as the Zeeman energy becomes larger than thermal and tunnel broadening \cite{FrolovPRL09}.  Second, a dip is observed between 5 and 8 T in each device, resulting from faster spin relaxation at the BSR condition as is described below.  The centers of the BSR dips in the four panels of Fig. \ref{by} vary due to changes in the 2DEG density (and hence $v_F$)  for different cooldowns.

Measurement of BSR in two orthogonal channels provides direct access to spin-orbit anisotropy. The channel orientation determines whether $[\bar{1}10]$ or $[110]$ component of momentum oscillates: \x-oriented channels induce oscillating $k_{\bar{1}10}$, while \y-oriented channels induce oscillating $k_{110}$. According to Eq. \ref{eq:bso}, an oscillating $k_{\bar{1}10}$ generates an oscillating $B^{so}_{110} \propto (\alpha-\beta)$, whereas $k_{110}$ generates $B^{so}_{\bar{1}10} \propto (\alpha+\beta)$ [Fig. \ref{device}(f)].

The spin signal vanishes almost completely inside the dip in \x\ channels,  while only a weak suppression is observed in \y\ channels [Fig.~\ref{by}]. At a qualitative level, this indicates that the oscillating spin-orbit field in \x\ channels is much larger than in \y\ channels, and based on the direction of anisotropy we conclude that $\alpha$ and $\beta$ as defined by Eq.~\ref{eq:bso} have opposite signs. BSR was measured in eight \x\ and three \y\ channels in this 2DEG. Details of the \vnl$(B^{ext})$ traces were different for each channel (even each cooldown) due to variations in QPC polarization, electron density, etc., but the dramatic difference in relative BSR dip depth for \x\ versus \y\ channels was consistent for all devices.

A quantitative determination of spin-orbit anisotropy is made by extracting spin relaxation length on resonance for the two channel orientations, $\lambda_{110}^{\rm{BSR}}$ and $\lambda_{\bar{1}10}^{\rm{BSR}}$. We first consider \x\ channels, used to determine $\lambda_{110}^{\rm{BSR}}$ [Figs.~\ref{by}(a) and \ref{by}(c)].  For the long channel, with $x_{id} = 25 ~\mu$m, the spin signal is suppressed below detectable levels between 4 and 6 T  [Fig.~\ref{by}(c)], indicating that $\lambda_{110}^{\rm{BSR}}\ll 25~\mu$m but making it difficult to determine an accurate numerical value.  (The shallow slope between 4 and 6 T is a slowly increasing thermoelectric signal, and not related to spin \cite{FrolovPRL09}.)  For the short channel, with $x_{id} = 5~\mu$m, the spin signal is detectable throughout the BSR dip, with a minimum at 6.5 T.  \vnl\ is similar before ($B^{ext}$ = 4 T) and after ($B^{ext}$ = 10 T)  the dip in Fig.~2(a), indicating that injector and the detector QPCs are fully polarized for $B^{ext} >$ 4 T in this device.  Thus changes in \vnl\ are due only to changes in spin relaxation length caused by BSR, and $\lambda$ can be deduced from solutions to a 1D spin diffusion equation, $\partial^2 V_{nl}/\partial x^2 = V_{nl}/\lambda^2$ giving $\lambda_{110}^{\rm{BSR}}=3.3\pm0.7~\mu$m \cite{suppmat, FrolovPRL09}.

Channels oriented along \y, on the other hand, showed a clear spin signal on resonance for both long and short geometries [Figs.~2(b) and 2(d)].   The signal did not saturate with field up to $B^{ext}$=12T in these devices, possibly due to QPC polarizations slowly increasing with field, so it was not possible to determine $\lambda_{\bar{1}10}^{\rm{BSR}}$ simply by comparing \vnl\ inside and outside the BSR dip.  Instead,  the $\Lambda$-gate in the long channel was used to change the channel length in situ [Fig.~1(d)]: a polarization-independent strategy for extracting $\lambda^{\rm{BSR}}$ \cite{FrolovPRL09}.  \vnl\ decreases when the $\Lambda$-gate is undepleted because the right-hand reservoir with equilibrium spin polarization is effectively brought closer to the detector. Spin relaxation length is extracted from the ratio \vnl($\Lambda$-gate on)/\vnl($\Lambda$-gate off) using the solution to the 1D diffusion equation, giving 65$\pm5~ \mu$m off resonance and 40$\pm5 ~\mu$m in the BSR dip \cite{suppmat}.

\begin{figure}[!h]
  \includegraphics[scale=1]{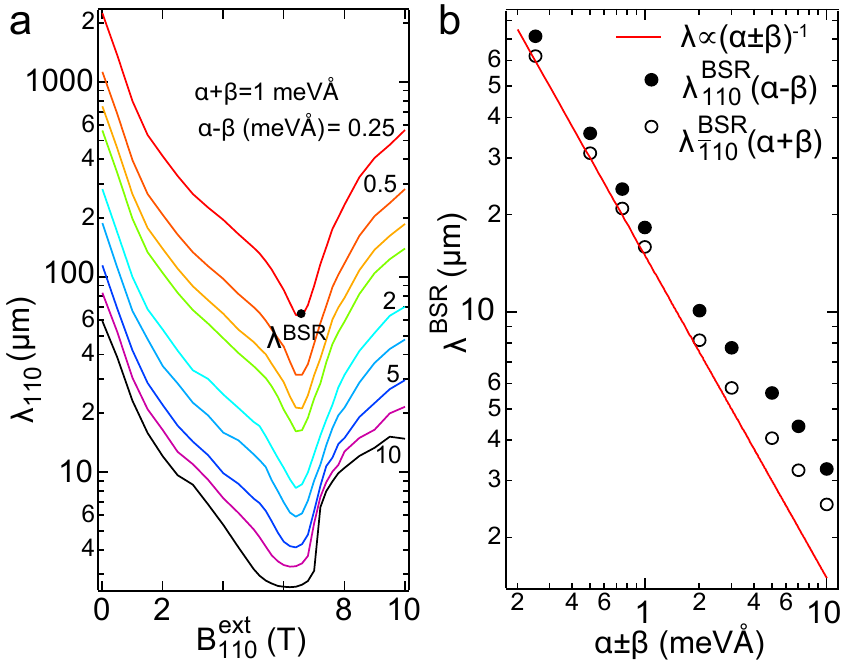}
  \caption{\textbf{(a)} Monte Carlo simulations of spin relaxation time, converted into spin relaxation length $\lambda$ (see text), using parameters $\ell=12~\mu$m, $v_F=8.4\times 10^4$ m/s, and $w=900$ nm corresponding to the channel in Fig.~2(a).  From top to bottom, curves represent ($\alpha - \beta$)=0.25, 0.5, 0.75, 1, 2, 3, 5, 7, and 10 meV\AA. The value of ($\alpha + \beta$) is fixed at 1 meV\AA. \textbf{(b)} Spin relaxation length at the center of the dip, $\lambda^{\rm{BSR}}$, for channel parameters in Fig.~2(a) ($\lambda^{\rm{BSR}}_{110}$) and Fig.~2(d) ($\lambda^{\rm{BSR}}_{\bar{1}10}$) plotted against the relevant combination of $\alpha$ and $\beta$. The relation $\lambda^{\rm{BSR}} \propto (\alpha\pm\beta)^{-1}$ valid in the weak spin-orbit limit is shown for reference (solid line).  For the channel in Fig.~2(d), $\ell=18~\mu$m, $v_F=9.9\times 10^4$ m/s, and $w=950$ nm.}
 	\label{sim}
\end{figure}

For Dyakonov-Perel' spin relaxation in which the spin precession length due to SOI is longer than the channel width, a spin relaxation time of approximately $\tau_{sr} \propto (B^{so})^{-2}$ is expected, corresponding to a spin relaxation length $\lambda=\sqrt{\tau_{sr} \ell v_F/2}\propto (B^{so})^{-1}$ where $\ell$ is the mean free path \cite{Luescher10, Kiselev:2000}.
In order to obtain more accurate estimates of $\tau_{sr}$, spin dynamics were simulated using Monte-Carlo techniques, averaging over an ensemble of realistic random trajectories based on the mean free path, $\ell$, channel width, $w$, and $v_F$ for a particular channel \cite{suppmat}.  The simulations provided estimates of $\tau_{sr}(B^{so})$ for the two channels, which were converted into $\lambda$ and matched to the experimental data.

An example is shown in Fig.~3(a), where $\lambda_{110}(B^{ext}_{\bar{1}10})$ is obtained for a wide range of spin-orbit parameters $(\alpha-\beta)$, from simulations based on $\ell, v_F$, and $w$ for the channel in Fig.~2(a).   The simulations confirmed that BSR in the 110 channel was sensitive only to $(\alpha-\beta)$, but nearly independent of $(\alpha+\beta)$, and vice versa (c.f. Eq. \ref{eq:bso}); the sum and difference of $\alpha$ and $\beta$ could therefore be fitted independently to the two channels, greatly facilitating the determination of both parameters.  An analogous simulation was performed for the channel in Fig.~2(d).  Based on this analysis, the values of $\lambda^{\rm{BSR}}$ found earlier gave ($\alpha-\beta$) = $7\pm2$ meV\AA\ and ($\alpha+\beta$) = $0.45\pm0.05$ meV\AA: an anisotropy of $|\alpha-\beta|/|\alpha+\beta|=15\pm5$.

\begin{figure}[!h]
  \includegraphics[scale=1]{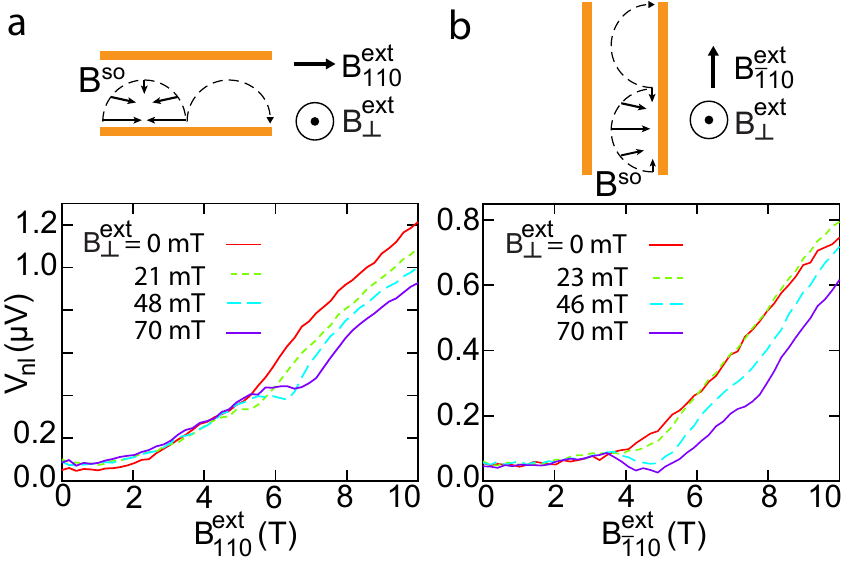}
  \caption{Ballistic spin resonance induced by out-of-plane magnetic field $B^{ext}_{\bot}$ in a \x\ channel, \textbf{(a)}, and a \y\ channel, \textbf{(b)}, both with $x_{id} = 25~\mu$m. Schematics on top show the channel and magnetic field orientations, and simplified cyclotron trajectories for the two panels.  In each case $V_{nl}$ in the dip must be compared to signal for $B^{ext}_{\bot}=0$, where no dip is seen,  to take into account partial QPC polarization.}
 	\label{bx}
\end{figure}

BSR can also be induced with $B^{ext}$ oriented primarily along the channel axis if the component of $B^{ext}$ perpendicular to the 2DEG plane, $B^{ext}_{\bot}$, is not zero \cite{FrolovNature09}. Electrons follow cyclotron trajectories due to $B^{ext}_{\bot}$, along which both $k_{\bar{1}10}$ and $k_{110}$ oscillate. This ensures an oscillating component of $B^{so}$ perpendicular to $B^{ext}$, as required for BSR. Figure \ref{bx} demonstrates how BSR develops with $B^{ext}_{\bot}$ in orthogonal channels. For $B^{ext}_{\bot} = 0$ the BSR dip is not observed because  the only oscillating component of the momentum is perpendicular to the channel.  The oscillating component of the spin-orbit field therefore lies along the channel axis, parallel to $B^{ext}$. At higher $B^{ext}_{\bot}$ the BSR dip grows because the component of momentum along the channel axis also begins to oscillate. The dip shifts to higher in-plane field with $B^{ext}_{\bot}$, as the bouncing frequency increases with decreasing cyclotron radius.

BSR from $B^{ext}_{\bot}$ in \x\ and \y\ channels confirms the strong spin-orbit anisotropy reported above. The dip in the \x\ channel is now weaker than in \y\ channel, because $B^{ext}_{110}$ probes the cyclotron-induced oscillation in $k_{110}$, the weak spin-orbit direction. In the \y\ channel BSR leads to a suppression of $V_{nl}$  by nearly 85\% compared to the signal for  $B^{ext}_{\bot}=0$ [Fig.~4(b)], while \vnl\ is only suppressed by 35\% in the \x\ channel.

In conclusion, BSR is used to estimate the degree of spin-orbit anisotropy in a GaAs/AlGaAs 2DEG. The spin relaxation length at BSR is found to be more than an order of magnitude shorter in channels oriented in the \x\ direction compared to  \y.  The extracted values for $\alpha$ and $\beta$ are within 20\% of each other, suggesting that in the future it may be possible to reach the regime $|\alpha|=|\beta|$ by tuning the 2DEG with a top gate \cite{KainzPRB03, MillerPRL03}.

Work at UBC supported by NSERC, CFI, and CIFAR.  W.W. acknowledges financial support by the Deutsche Forschungsgemeinschaft (DFG) in the framework of the program ``Halbleiter-Spintronik" (SPP 1285).

\end{document}